\def\rfr#1{eq. (\ref{#1})}

\def\bm#1{{\mbox{\boldmath$#1$\unboldmath}}}
\def\asec{$''$ cy$^{-1}$}

\def\dert#1#2{\frac{{{d}}{#1}}{{{d}}{#2}}}              

\def\asec{$''$ cy$^{-1}$}

\def\bar{\begin{eqnarray}}
\def\ear{\end{eqnarray}}
\def\bb{\bibitem}
\def\eqi{\begin{equation}}
\def\eqf{\end{equation}}
\def\eqI{\begin{equation}}
\def\eqF{\end{equation}}
\def\eqIa{\begin{eqnarray}}
\def\eqFa{\end{eqnarray}}
\def\rp#1#2{{#1\over#2}}

\def\lb#1{\label{#1}}
\def\beq{\begin{equation}}
\def\eeq{\end{equation}}
\def\bcal#1{\bm{\mathcal{{#1}}}}

\def\oc2{$\mathcal{O}(c^{-2})$}

\def\bds#1{\boldsymbol{#1}}

\documentclass{aastex}
\usepackage{amsmath,amsthm,amscd,amssymb}
\usepackage{latexsym}
\usepackage{graphicx,epsfig}

\RequirePackage{color}

\begin{document}

\title{Constraining Post-Newtonian $f(R)$
Gravity in the Solar System}

\shorttitle{Constraining $f(R)$
Gravity in the Solar System} \shortauthors{M.L. Ruggiero and L. Iorio}

\author{Matteo Luca Ruggiero}
\affil{Dipartimento di Fisica del Politecnico di Torino and
INFN-Sezione di Torino, Corso Duca degli Abruzzi 24, 10129, Torino
(TO), Italy. E-mail: matteo.ruggiero@polito.it} \and
\author{Lorenzo Iorio }
\affil{INFN-Sezione di Pisa. Permanent address for correspondence:
Viale Unit\`{a} di Italia 68, 70125, Bari (BA), Italy. E-mail:
lorenzo.iorio@libero.it}

\begin{abstract}
We consider some models of $f(R)$ gravity that can be used to describe, in a suitable weak-field limit, the gravitational field of the Sun. Using a perturbative approach, we focus on the impact that the  modifications
 of the gravitational field, due to the non-linearity of the gravity Lagrangian, have on the
Solar System dynamics. We compare the theoretical predictions for the precession of the longitude of the pericentre $\varpi$ of a test particle with the corrections to the standard Newtonian-Einsteinian precessions of the longitudes of perihelia of some planets of the Solar System recently estimated by E.V. Pitjeva by fitting large data sets with various versions of the EPM
ephemerides.
\end{abstract}


\section{Introduction}\lb{intro}

General Relativity (GR) has passed with excellent results many
observational tests: a satisfactory agreement comes both from
Solar System tests and from binary pulsars
observations. As a matter of fact  (see e.g. \citep{Will06}), the
current values of the PPN parameters are in agreement with GR
predictions and, consequently, Einstein's theory is the classical theory of gravitational interactions
accepted nowadays.

However, observations seem to question the general relativistic model of gravitational
interactions on large scales.  On the one hand, the data coming
from the rotation curves of spiral galaxies
\citep{Binney87} cannot be explained on the basis of Newtonian
gravity or GR: the existence of a peculiar form of matter is
postulated to reconcile the theoretical model with observations,
i.e. dark matter, which is supposed to be a cold and pressureless
medium, whose distribution is that of a spherical halo around the
galaxies.
Furthermore, dark matter can explain the mass
discrepancy in galactic clusters \citep{Clowe06}. On the other
hand, a lot of observations, such as the light curves of the type
Ia supernov{\ae} and the cosmic microwave background (CMB)
experiments \citep{Riess98,Perlmutter99,Bennet03}, firmly state
that our Universe is now undergoing a phase of accelerated
expansion. Actually, the present acceleration of the Universe
cannot be explained, within GR, unless the existence of a cosmic
fluid having exotic properties is postulate, i.e. dark energy or introducing a  cosmological constant which, in turn, brings
about other problems, concerning its nature and origin
\citep{peebles03}.

The main problem one has to face with dark matter and dark energy (or the cosmological constant)
is understanding their nature, since they are introduced as
\textit{ad hoc} gravity sources  in GR or its weak-field limit,
Newtonian gravity.

In order to explain the observations  another possibility exists:
the query for dark matter and dark energy points out the failure
of GR (and its approximation, Newtonian gravity) to deal with
gravitational interaction at galactic, intergalactic and
cosmological scales. The latter viewpoint led to the introduction
of various modified gravity models.

In this paper, we are concerned with the  so called $f(R)$
theories of gravity, where the gravitational Lagrangian depends on
a function $f$ of the scalar curvature $R$
(see  \citep{Capofranc07,sotfar08} and references therein). These theories
are also referred to as ``extended theories of gravity'', since
they naturally generalize GR: in fact, when $f(R)=R$ the action reduces to the usual
Einstein-Hilbert action, and Einstein's theory is obtained. These theories can be studied in the metric
formalism, where the action is varied with respect to metric
tensor, and in the Palatini formalism, where the action is varied
with respect to the metric and the affine connection, which are
supposed to be independent from one another (actually, there is
also the possibility that the matter part of the action depends on
the affine connection, and is then  varied with respect to it:
this is the so-called metric-affine formalism, but we are not
concerned with this approach in this paper). In general, the two
approaches are not equivalent:  the solutions of the Palatini
field equations are a subset on solutions of the metric field
equations \citep{magnano}.

Actually, $f(R)$ theories  provide cosmologically viable models, where both
the inflation phase and the accelerated expansion are reproduced
(see \citep{odi07,odi08a,odi08b} and references therein). Furthermore, they have been used to explain the
rotation curves of galaxies without need for dark matter \citep{Capo07,Martins2007}.

However, because of the excellent agreement of GR with  Solar
System and binary pulsar observations,  every theory that aims at
explaining galaxies dynamics and the accelerated expansion of the
Universe, should reproduce GR at the Solar System scale, i.e. in a
suitable weak-field limit. In other words, also for $f(R)$
theories the constraint holds to have correct Newtonian and
post-Newtonian limits. This issue has been lively debated in the
recent literature, where different approaches to the problem have
been taken into account, both in the Palatini and metric formalism.
A thorough discussion  can be found in the recent review by \citet{sotfar08}.
In summary, with respect to the weak-field tests and, more in general, the non cosmological solutions (see e.g. \citep{faraoni08a}),  it seems that there are difficulties in considering  Palatini $f(R)$ gravity as a viable theory because the Cauchy problem is ill-posed and, furthermore,  curvature singularities arise when dialing with simple stellar models; as for metric
$f(R)$ gravity, there are models that are in agreement with the weak-field tests, but  it seems that curvature singularities exist, in this case, for compact relativistic stars.

Without going into the details of this interesting debate, in this paper we want to test some models of $f(R)$ gravity that can be used to describe the gravitational field of the Sun.
In particular, we are going to  examine the impact
that the modifications of the  gravitational field of GR have on the
Solar System dynamics. We apply a perturbative approach to
compare the $f(R)$-induced  secular precession of the longitude of the pericentre $\varpi$ of a test particle with
the latest determinations of the corrections to the usual perihelion precessions coming from fits of huge planetary data sets with various versions of the EPM 
ephemerides \citep{Pit05a,Pit05b,Pit06,PIT07,Pit08}.

The paper is organized as follows: in Section \ref{sec:theof} we briefly review the
theoretical formalism of $f(R)$ gravity, both in the metric and Palatini approach,
then, in Section \ref{sec:corr} we outline a general approach to the perturbations of the gravitational field of GR, due to the non-linearity of the gravity Lagrangian.
In Section \ref{sec:fRsol} we compare the theoretical predictions with the observations. Finally, discussion and
conclusions are in Section \ref{sec:disconc}.

\section{The field equations of $f(R)$ gravity} \label{sec:theof}

In this Section, we introduce the
field equations of $f(R)$ gravity. We shall consider both the metric and the
Palatini approach (see, e.g., \citep{Capofranc07} and \citep{sotfar08}).

The equations of motion of  $f(R)$ extended theories of gravity
can be obtained by a variational principle, starting from the
action:
\begin{equation}
A=A_{\mathrm{grav}}+A_{\mathrm{mat}}=\int [ \sqrt{g} f (R)+2\chi
L_{\mathrm{mat}} (\psi, \nabla \psi) ]  \; d^{4}x.
\label{eq:actionf(R)}
\end{equation}
The gravitational part of the Lagrangian is represented by a
function $f (R)$ of  the scalar curvature $R$. The total
Lagrangian contains also a first order matter part
$L_{\mathrm{mat}}$, functionally depending on matter fields $\Psi$,
together with their first derivatives, equipped with a
gravitational coupling constant $\chi=\frac{8\pi G}{c^4}$. In the
metric formalism, $\Gamma$ is supposed to
be the Levi-Civita connection of $g$ and, consequently, the scalar
curvature $R$ has to be intended as $R\equiv R(g)
=g^{\alpha\beta}R_{\alpha \beta}(g)$. On the contrary, in the
Palatini formalism the metric $g$ and the affine connection
$\Gamma$ are supposed to be independent, so that the scalar
curvature $R$ has to be intended as $R\equiv R( g,\Gamma)
=g^{\alpha\beta}R_{\alpha \beta}(\Gamma )$, where $R_{\mu \nu
}(\Gamma )$ is the Ricci-like tensor of the connection $\Gamma$.

In the metric formalism the action (\ref{eq:actionf(R)})
is varied with respect to the metric $g$, and one obtains the
following field equations
\begin{align}
f'(R) R_{\mu \nu }-\frac{1}{2}f(R)g_{\mu \nu }-\left(\nabla _{\mu
}\nabla _{\nu }-g_{\mu \nu }\square \right)f'(R)= \frac{8\pi
G}{c^4} \,T_{\mu \nu}, \label{eq:fieldmetric1}
\end{align}
where $f'(R)=df(R)/d R$, and  $T^{\mu\nu}=-\frac{2}{\sqrt
g}\frac{\delta L_{\mathrm{mat}}}{\delta g_{\mu\nu}}$ is the
standard minimally coupled matter energy-momentum tensor. The
contraction of  the field equations (\ref{eq:fieldmetric1}) with
the metric tensor leads to the scalar equation
\begin{align}  \label{eq:fieldmetricscalar1}
3\square f'(R)+f'(R)R-2f(R)=\frac{8\pi G}{c^4} T,
\end{align}
where $T$ is the trace of the energy-momentum tensor. Eq.
(\ref{eq:fieldmetricscalar1}) is a differential equation for the
scalar curvature $R$.

In the Palatini formalism, by
independent variations with respect to the metric $g$ and the
connection $\Gamma$, we obtain the following equations of motion:
\begin{eqnarray}
f^{\prime }(R) R_{(\mu\nu)}(\Gamma)-\frac{1}{2} f(R)  g_{\mu \nu
}&=&\frac{8\pi G}{c^4} T_{\mu \nu },  \label{ffv1}\\
\nabla _{\alpha }^{\Gamma }[ \sqrt{g} f^\prime (R) g^{\mu \nu
})&=&0, \label{ffv2}
\end{eqnarray}
where  $\nabla^{\Gamma}$ means covariant derivative with respect
to the connection $\Gamma$.  Actually, it is possible to show
\citep{FFVa,FFVb} that the manifold $M$, which is the model of the
space-time, can be a posteriori endowed with a bi-metric structure
$(M,g,h)$  equivalent to the original metric-affine structure
$(M,g,\Gamma)$, where $\Gamma$ is assumed to be the Levi-Civita
connection of $h$. The two metrics are conformally related by
\begin{equation}\label{h_met2}
h_{\mu \nu }=f^\prime (R)  \;  g_{\mu \nu }.
\end{equation}
The equation of motion (\ref{ffv1})
can be supplemented by the scalar-valued equation obtained by
taking the contraction of (\ref{ffv1}) with the metric tensor:
\begin{equation}
f^{\prime} (R) R-2 f(R)= \frac{8\pi G}{c^4} T.  \label{ss}
\end{equation}
Equation (\ref{ss}) is an algebraic equation for the scalar
curvature $R$.

In order to compare the predictions of $f(R)$ gravity with Solar System data, we have to consider the solutions of the field
equations (\ref{eq:fieldmetric1}),(\ref{ffv1}),(\ref{ffv2}) -
supplemented by the constraints
(\ref{eq:fieldmetricscalar1}),(\ref{ss}) - in vacuum, since tests are based on the observations of the dynamics of the planets in the gravitational field of the
Sun.

In particular, the vacuum field equations in the metric approach
read
\begin{align}
f'(R) R_{\mu \nu }-\frac{1}{2}f(R)g_{\mu \nu }-\left(\nabla _{\mu
}\nabla _{\nu }-g_{\mu \nu }\square \right)f'(R)= 0,
\label{eq:fieldmetric1vac}
\end{align}
supplemented with the scalar equation
\begin{align}  \label{eq:fieldmetricscalar1vac}
3\square f'(R)+f'(R)R-2f(R)=0.
\end{align}
In the Palatini approach, the field equations become
\begin{eqnarray}
f^{\prime }(R) R_{(\mu\nu)}(\Gamma)-\frac{1}{2} f(R)  g_{\mu \nu
}&=&0,  \label{ffv1vac}\\
\nabla _{\alpha }^{\Gamma }[ \sqrt{g} f^\prime (R) g^{\mu \nu
})&=&0, \label{ffv2vac}
\end{eqnarray}
and they are supplemented by the scalar equation
\begin{equation}
f^{\prime} (R) R-2 f(R)=0  \label{ssvac}
\end{equation}
We want to point out some general features of the scalar equations (\ref{eq:fieldmetricscalar1vac}) and (\ref{ssvac}), which
can help to understand the differences between the vacuum solutions in the two formalisms.

In  Palatini $f(R)$ gravity,  the trace equation (\ref{ssvac}) is
an algebraic equation for $R$, which admits constant solutions
$R=c_{i}$ \citep{FFVa}, and it is identically satisfied if $f(R)$
is proportional to $R^2$. As a consequence, it is easy to verify
that (if $f'(R) \neq 0$) the field equations become
\begin{equation}
R_{\mu\nu}=\frac 1 4 R g_{\mu\nu}
\label{eq:gralambda1}
\end{equation}
which are the same as GR field equations with a cosmological
constant. In other words, in the Palatini formalism, in vacuum, we can
have only solutions that describe space-times with constant scalar
curvature $R$. Summarizing, eq. (\ref{eq:gralambda1})
suggests that all GR solutions with cosmological constant  are
solutions of vacuum Palatini field equations: the function $f(R)$
determines the  solutions of algebraic equation (\ref{ssvac}).

In metric $f(R)$ gravity the trace equation
(\ref{eq:fieldmetricscalar1vac}) is a differential equation for
$R$: this means that, in general, it admits more solutions than the
corresponding Palatini equation. In particular, we notice that if
$R=\mathrm{constant}$ we obtain the Palatini case: so for a given
$f(R)$ function, in vacuum,  the solutions of the field
equations of Palatini $f(R)$ gravity are a subset of the solutions
of the field equations of metric $f(R)$ gravity \citep{magnano}; however, in metric $f(R)$ gravity, vacuum solutions  with variable
$R$ are allowed too (see, e.g., \citep{metricspherically}).

\section{Corrections to the gravitational potential} \label{sec:corr}

We have seen in the previous Section that, when $f(R) \neq R$, the field equations of $f(R)$ gravity are different from those of GR. Thus, it is evident that
the solutions of such modified field equations describing the gravitational field of a point-like mass (e.g. the Sun) contain corrections to the GR solutions, both at
Newtonian and post-Newtonian level. However, these corrections have to be small enough not to contradict the known tests of GR. Thus, it is possible to treat them perturbatively  to evaluate their impact on the dynamics of the Solar System planets.

In this Section we want to outline the general procedure that we are going to apply to some solutions of $f(R)$ gravity that can be used to describe the gravitational field of the Sun, in order to compare the predictions of these gravity models with the existing data.

In general, we are going to deal with spherically symmetrical metrics, describing the space-time around a point-like mass $M$, which can be endowed with proper angular momentum $\bds{J}$. The weak-field and slow-motion approximations of these metrics will be sufficient for our purposes. Generally speaking, this means that the deviations from GR will be linear in some parameters deriving from the specific $f(R)$ gravity model.

On using spherical isotropic coordinates, these metric have the general form\footnote{If not otherwise stated, here and henceforth we use units
such that $G=c=1$.}
\begin{equation} ds^2=A(r)
dt^2+B(r)\left( dr^2+r^2d\vartheta^2+r^2\sin^2 \vartheta
d\varphi^2\right)+2C(r)sin^2 \theta dt d\varphi,
\label{eq:metricasferica1}
\end{equation}
where the angular momentum $\bds{J}$ is assumed to be perpendicular to the $\theta=\pi/2$ plane.

The gravitational (scalar) potential $\Phi(r)$ is read from
the $A(r)$  function
\beq  A(r)=1+2\Phi(r). \label{eq:Arphir1} \eeq
According to what stated before, we expect a gravitational potential in the form
\beq
\Phi(r)=\Phi^{\rm N}(r)+\Delta \Phi(r), \label{eq:deltaphidef1} \eeq
where $\Phi^{\rm N}(r)=-\frac{M}{r}$ is the Newtonian potential of a point-like mass $M$, and
$\Delta \Phi(r) \ll \Phi_{N}(r)$ is a correction vanishing for $f(R) \rightarrow R$.\\

The $C(r)$ function accounts for the presence of the so-called
gravito-magnetic effects \citep{ruggiero02,mashhoon03} induced by the rotation of the source of the
gravitational field. In GR $C(r)$ is given by the suitable component of the gravito-magnetic vector potential of a gravito-magnetic dipole, i.e. $A^{\rm GR}_{\varphi}(r)=-\frac{2J}{r}$ (see e.g. \citep{mashhoon03}). As a consequence, we expect that the $C(r)$ function has the form
\beq C(r)=A^{\rm GR}_{\varphi}(r)+\Delta A_{\varphi}(r) \label{eq:deltaAdef1} \eeq
where, again, $\Delta A_{\varphi}(r) \ll  A_{\varphi}(r)^{\rm GR}$ is a correction vanishing for $f(R) \rightarrow R$.

We can use the gravito-electromagnetic \citep{ruggiero02,mashhoon03} formalism to describe the total perturbing acceleration felt by a test particle  in the metric (\ref{eq:metricasferica1})
\beq  \bds W=-\bcal{E}^{\rm G}-2{\bds{v}}{}\bds\times\bcal{B}^{\rm G} \label{eq:acclorentz1}\eeq
where
\beq \bcal{E}^{\rm G}=-\frac{d \Delta \Phi(r)}{dr} \bds{\hat r} \label{eq:defEg1}\eeq
and
\beq \bcal{B}^{\rm G}=\bds{\nabla} \bds\times\bds{A}, \quad \bds{A}=\frac{\Delta A_{\varphi}}{r \sin\theta } \label{eq:defBg1}\eeq

Hence, given the perturbing acceleration (\ref{eq:acclorentz1}), we can calculate its effects on planetary motions within standard perturbative
schemes (see, e.g., \citet{roy}). We may use the Gauss equations for the variations of the elements, which enable us to study the perturbations of the Keplerian orbital elements due to a generic perturbing acceleration, whatever its physical origin is.
The Gauss equations for the
variations of the semi-major axis $a$, the eccentricity $e$, the
inclination $i$, the longitude of the ascending node $\Omega$, the
argument of pericentre $\omega$ and the mean anomaly $\mathcal{M}$
of a test particle in the gravitational field of a body $M$ are  \citep{roy}

\bar
\dert{a}{t} & = &  \rp{2}{  {\overline{n}} \sqrt{1-e^2}  }    \left[eW_r\sin v+W_{\tau}\left(\rp{p}{r}\right)\right],\lb{smax}\\
\dert{e}{t} & = & \rp{\sqrt{1-e^2}}{{\overline{n}}a}\left\{W_r\sin v+W_{\tau}\left[\cos v+\rp{1}{e}\left(1-\rp{r}{a}\right)\right]\right\},\\
\dert{i}{t} & = & \rp{1}{{\overline{n}}a\sqrt{1-e^2}}\ W_{\nu}\left(\rp{r}{a}\right)\cos (\omega+v),\lb{in}\\
\dert{\Omega}{t} & = & \rp{1}{{\overline{n}}a\sin i\sqrt{1-e^2}}\ W_{\nu}\left(\rp{r}{a}\right)\sin (\omega+v),\lb{nod}\\
\dert{\omega}{t} & = & -\cos i\dert{\Omega}{t}+\rp{\sqrt{1-e^2}}{{\overline{n}}ae}\left[-W_r\cos v+W_{\tau}\left(1+\rp{r}{p}\right)\sin v\right],\lb{perigeo}\\
\dert{\mathcal{M}}{t} & = & {\overline{n}} -\rp{2}{{\overline{n}}a}\
W_r\left(\rp{r}{a}\right)-\sqrt{1-e^2}\left(\dert{\omega}{t}+\cos
i\dert{\Omega}{t}\right),\lb{manom} \ear

in which ${\overline{n}}=2\pi/P$ is the  mean motion\footnote{For an
unperturbed Keplerian ellipse it is ${\overline{n}}=\sqrt{GM/a^3}$.},
$P$ is the test particle's orbital period, $v$ is the true
anomaly counted from the pericentre, $p=a(1-e^2)$ is the semilatus
rectum of the Keplerian ellipse, $W_r,\ W_{\tau},\ W_{\nu}$ are the
radial, transverse (in-plane components) and  normal (out-of-plane
component) projections of the perturbing acceleration
$\bds{W}$, respectively, on the orthonormal frame
$\{\boldsymbol{\hat{r}},\boldsymbol{\hat{\tau}},\boldsymbol{\hat{\nu}}\}$
comoving with the particle.

For our purposes it is useful to consider the longitude of the pericenter $\varpi=\omega+\cos i\ \Omega$.  The Gauss equation for its variation under the action of an
entirely radial perturbing acceleration $W_r$ is \eqi\dert\varpi t
=-\rp{\sqrt{1-e^2}}{{\overline{n}} ae}W_r\cos v.\lb{gaus}\eqf
After being
evaluated onto the unperturbed Keplerian ellipse, the acceleration
(\ref{eq:acclorentz1}) must be inserted into \rfr{gaus}; then, the average over one
orbital period $P$ must be performed. To this end
it is useful also to recall the following relations
where also the eccentric anomaly $E$ is used
\begin{equation}\left\{\begin{array}{lll}
r=a(1-e\cos E),\\\\
dt=\rp{(1-e\cos E)}{{\overline{n}}}dE
,\\\\
\cos v=\rp{\cos E-e}{1-e\cos E
},\\\\
\sin v=\rp{\sin E\sqrt{1-e^2}}{1-e\cos E}.
\lb{eccen}\end{array}\right.\end{equation}

In fact, what we aim at is evaluating the perturbations
induced on the longitudes of the perihelia by the corrections to
the gravitational field due to $f(R)$ gravity, in
order to compare them with the latest observational determinations. The
astronomer E.V. Pitjeva (Institute of Applied Astronomy, Russian Academy of Sciences, St. Petersburg) processed
almost one century of data of different types for the major bodies
of the Solar System to improve the EPM planetary ephemerides
\citep{Pit05a,PIT07,Pit08}. Among other things, she  simultaneously estimated
corrections to the secular rates of the longitudes of perihelia
$\varpi$ of the inner \citep{Pit05b} and of some of the outer
\citep{PIT07,Pit08} planets of the Solar System as fit-for parameters of
 global solutions in which she contrasted, in a least-square way,
the observations to their predicted values computed with a
complete set of dynamical force models including all the known
Newtonian (solar quadrupole mass moment $J_2$, $N-$body interactions with the major planets, 301 biggest asteroids, massive ring of the small asteroids, 20 largest trans-Neptunian objects and massive ring for the other ones) and Einsteinian\footnote{The general relativistic gravito-magnetic Lense-Thirring force  has not yet been modeled.} features of motion. As a consequence,
any force that is not present in Newtonian gravity or GR is, in
principle, accounted for by the estimated corrections to the usual apsidal
precessions.  For the sake of completeness, we reproduce in
tables \ref{tavola1} and  \ref{tavola2} the estimated perihelia
extra-precessions for inner and outer planets, respectively.

{\small\begin{table}\caption{ Inner planets. First row: estimated
perihelion extra-precessions, from Table 3 of \citep{Pit05b}. The
quoted errors are not the formal ones but are realistic. The units
are arc-seconds per century (\asec). Second row: semi-major axes, in
Astronomical Units (AU). Their formal errors are in Table IV of
\citep{Pit05a}, in m. Third row: eccentricities. Fourth row:
orbital periods in years.}\label{tavola1}

\begin{tabular}{cccc} \noalign{\hrule height 1.5pt}

& Mercury & Earth & Mars\\
\hline
$\left<\dot\varpi\right>$ (\asec) & $-0.0036\pm 0.0050$ & $-0.0002\pm 0.0004$ & $0.0001\pm 0.0005$\\
$a$ (AU) & 0.387 & 1.000 & 1.523 \\
$e$ & 0.2056 & 0.0167 & 0.0934\\
$P$ (yr) & 0.24 &  1.00 & 1.88\\
\hline

\noalign{\hrule height 1.5pt}
\end{tabular}

\end{table}}

{\small\begin{table}\caption{ Outer planets. First row: estimated
perihelion extra-precessions \citep{Pit06}. The quoted
uncertainties are the formal, statistical errors re-scaled by a
factor 10 in order to get the realistic ones.  The units are
arc-seconds per century (\asec). Second row: semi-major axes, in
Astronomical Units (AU). Their formal errors are in Table IV of
\citep{Pit05a}, in m. Third row: eccentricities. Fourth row:
orbital periods in years.}\label{tavola2}

\begin{tabular}{cccc} \noalign{\hrule height 1.5pt}

& Jupiter & Saturn & Uranus\\
\hline
 $\left<\dot\varpi\right>$ (\asec) & $0.0062\pm 0.036$ & $-0.92\pm 2.9$ & $0.57\pm 13.0$\\
$a$ (AU) & 5.203 & 9.537 & 19.191\\
$e$ &  0.0483 & 0.0541 & 0.0471 \\
$P$ (yr) & 11.86 & 29.45 & 84.07\\
\hline

\noalign{\hrule height 1.5pt}
\end{tabular}

\end{table}}


What we want to do is to see whether the estimated
perihelia extra-precessions are compatible with the
perturbations of the gravitational field deriving from the non-linearity of $f(R)$.

Now, let us briefly outline how we are going to put $f(R)$ gravity on the test. In general a correction to the gravitational field of GR due to the non linearity of the gravity Lagrangian, in the weak-field and slow motion approximation, can be parameterized in terms of a parameter $\kappa$, where $\kappa \rightarrow 0$ as far as $f(R) \rightarrow R$. In other words, $\kappa$ is a measure of the non-linearity of the Lagrangian. Let ${\mathcal{P}}({f(R)})$ be the prediction of a certain effect induced by these modified gravity models,
e.g. the secular precession of the perihelion of a planet: for all the $f(R)$ models that we are going to consider below, it turns out that
 \eqi{\mathcal{P}}({f(R)}) = \kappa g(a,e), \label{eq:mod11}\eqf
where $g$ is a function of the system's
orbital parameters $a$ (semi-major axis) and $e$ (eccentricity);
such $g$ is a peculiar consequence of the $f(R)$ gravity model. Now, let us take the ratio of ${\mathcal{P}}({f(R)})$ for two different systems {\rm A} and {\rm B}, e.g. two
Solar System's planets: ${\mathcal{P}}_{\rm A}({f(R)})/{\mathcal{P}}_{\rm B}({f(R)}) = g_{\rm A}/g_{\rm B}$.
The model's parameter $\kappa$ has now been canceled, but we still have
a prediction that retains a peculiar signature of that model, i.e.
$g_{\rm A}/g_{\rm B}$. Of course, such a prediction is valid if we
assume $\kappa$ is not zero, which is just the case both theoretically
(only if $f(R)=R$ then $\kappa=0$) and observationally because $\kappa$ is
usually determined by other independent long-range
astrophysical/cosmological observations. Otherwise, one would have
the meaningless prediction $0/0$. The case $\kappa=0$ (or
$\kappa\leq\overline{\kappa}$, i.e. when $\kappa$ is negligibly small) can be, instead, usually tested  by taking
one perihelion precession at a time. If we have observational
determinations ${\mathcal{O}}$ for {\rm A} and {\rm B} of the
effect considered above  such that they are affected
also\footnote{If they are differential quantities constructed by
contrasting observations to predictions obtained by analytical
force models of canonical Newtonian/Einsteinian effects,
${\mathcal{O}}$ are, in principle, affected also by the
mis-modeling in them.} by  the ${f(R)}$ gravity model (it is just the case for
the purely phenomenologically estimated corrections to the
standard Newton-Einstein perihelion precessions, since any ${f(R)}$ gravity model has not been included in the  dynamical force models of
the ephemerides adjusted to the planetary data in the least-square
parameters' estimation process by Pitjeva \citep{Pit05a,Pit05b}),
we can construct ${\mathcal{O}}_{\rm A}/\mathcal{O}_{\rm B}$ and
compare it with the prediction for it by ${f(R)}$, i.e. with
$g_{\rm A}/g_{\rm B}$. Note that
$\delta{\mathcal{O}}/{\mathcal{O}}>1$ only means that
${\mathcal{O}}$ is compatible with zero, being possible a nonzero
value smaller than $\delta{\mathcal{O}}$. Thus, it is perfectly
meaningful to construct ${\mathcal{O}}_{\rm A}/\mathcal{O}_{\rm
B}$. Its uncertainty will be conservatively evaluated as
$|1/{\mathcal{O}}_{\rm B}|\delta{\mathcal{O}}_{\rm A} +
|{\mathcal{O}}_{\rm A}/{\mathcal{O}}_{\rm
B}^2|\delta{\mathcal{O}}_{\rm B}$. As a result,
${\mathcal{O}}_{\rm A}/\mathcal{O}_{\rm B}$ will be compatible
with zero. Now, the question is: Is it the same for $g_{\rm
A}/g_{\rm B}$ as well? If yes, i.e. if \eqi\rp{{\mathcal{O}}_{\rm
A}}{\mathcal{O}_{\rm B}}=\rp{{\mathcal{P}}_{\rm A}({ f(R)})}{{\mathcal{P}}_{\rm B}({f(R)})}\eqf within the
errors, or, equivalently, if \eqi\left|\rp{{\mathcal{O}}_{\rm
A}}{\mathcal{O}_{\rm B}} - \rp{{\mathcal{P}}_{\rm A}({
f(R)})}{{\mathcal{P}}_{\rm B}({ f(R)})}\right|=0\eqf within
the errors,  the $f(R)$ gravity model examined can still be considered compatible with the data, otherwise it is seriously challenged.

In next Section some solutions of $f(R)$ gravity that can be used to describe the gravitational field of the Sun will be tested according to the procedure that we have just described.

\section{$f(R)$ weak-field solutions} \label{sec:fRsol}

In this Section we introduce some solutions of $f(R)$ gravity that have been used in the literature to the describe the weak gravitational field, and that
can be considered as suitable models of the gravitational field of the Sun. We consider these modified gravitational fields and, within the perturbative scheme outlined above,  compare the theoretical predictions with the estimated extra-precessions of the planetary perihelia.

\subsection{Power law corrections } \label{ssec:powerlaw1}

Starting from a Lagrangian of the form $f(R)=f_0 R^n$, \citet{Capo07}, in the metric approach, look for
solutions describing the gravitational field of a point-like
source, in order to reproduce the galaxies rotation curves without need for dark matter.
 As a result, they obtain the following power-law form for
the gravitational potential:

\begin{equation}
\Phi(r)=-\frac{M}{r}\left[1+\left(\frac{r}{r_c}
\right)^\beta\right]. \label{eq:cappowlaw1}
\end{equation}

The deviation from the Newtonian potential  is parameterized by a
power law, with two free parameters $\beta$ and $r_c$. In
particular, $\beta$ is related to $n$, i.e. the exponential of the
Ricci scalar in $f(R)=f_0 R^n$.

In this case, the correction to the gravitational potential is
\eqi \Delta
\Phi(r)=-\rp{M}{r}\left(\rp{r}{r_c}\right)^{\beta},\lb{capopot}\eqf
which clearly leads to the radial acceleration \eqi W_r=
\rp{(\beta -1)M}{r_c^{\beta}}r^{\beta-2} \ \lb{accelcapo}\eqf

It yields the following perihelion precession  \citep{SYR}
\eqi\left<\dot\varpi\right>=\rp{(\beta-1)\sqrt{M}}{2\pi r_c^{\beta}}a^{\beta-\rp{3}{2}}G(e; \beta),\lb{capozz_A}\eqf
with $G(e_{\rm A};\beta)/G(e_{\rm B};\beta)\approx 1$ for all the planets of the Solar System.

\citet{Capo07} find $\beta = 0.817$ from a successful fit of several galactic rotation curves with no dark matter; it is ruled out by comparing for several pairs of planets $\Delta\dot\varpi_{\rm A}/\Delta\dot\varpi_{\rm B}$ to $\mathcal{P}_{\rm A}/\mathcal{P}_{\rm B}$ obtained from \rfr{capozz_A} \citep{SYR}.

\subsection{Schwarzschild-de
Sitter-like corrections} \label{ssec:pert}

The field equations (\ref{ffv1}-\ref{ffv2}) and the structural
equation (\ref{ss}), in the Palatini formalism have the
spherically symmetrical solution (see \citep{allemandi05} and \citep{ruggiero06}):
\begin{equation}
ds^2=\left(1-\frac{2M}{r}-\frac{k }{3}r^2\right)dt^2 -
\left(1+\frac{2M}{r} -\frac{k }{6}r^2
\right)\left(dr^2+r^2 d\theta^2+r^2 \sin^2 \theta d\phi^2\right), \label{eq:metricasds}
\end{equation}
$k=c_i/4$, where $R=c_i$ is any of the solutions of the
structural equation (\ref{ss}).
In this case, we may write \beq \Phi(r)=-\frac{M}{r}+\frac{k
r^2}{6}. \label{eq:sdspot1}\eeq
In this case, the perturbing potential is \beq \Delta
\Phi(r)=\frac{kr^2}{6}, \label{eq:pertsds1} \eeq   and the induced
the perturbing acceleration is \eqi W_r=-\rp{1}{3} k r.\lb{acc_des}\eqf

As any other Hooke-type extra-acceleration, \rfr{acc_des} induces a secular perihelion precession  \citep{kerr03,AdA}
\eqi\left\langle\dot\varpi\right\rangle\propto \rp{k}{{\overline{n}}}=k\sqrt{\rp{a^3}{M}}.\lb{hooke}\eqf
By using \rfr{hooke} to construct $\mathcal{P}_{\rm A}/\mathcal{P}_{\rm B}$ for different pairs of Solar System's planets
and comparing them  to $\Delta\dot\varpi_{\rm A}/\Delta\dot\varpi_{\rm B}$  yield a negative answer \citep{AdA}.

\subsection{Logarithmic corrections} \label{ssec:logs}

\citet{Sob07}  aims at determining a
$f(R)$ able to explain the rotation curves of the galaxies obtained. In particular, working in the metric approach, solutions with $R$ variable with the radial coordinate $r$ are obtained. In this context, the gravitational potential reads:

\beq \Phi(R) =-\frac{M}{r}
+\frac{\alpha}{2}+\frac{\alpha}{2}\ln(r/2M). \label{eq:sobo1}
\eeq

The parameter $\alpha$ can be related to Modified Newtonian
Dynamics (MOND, see e.g. \citep{Mil83}) characteristic
acceleration $A_0$.

The Logarithmic correction to the Newtonian gravitational
potential assumes the form \eqi \Delta \Phi(r)=-\gamma M \ln
\left(\rp{r}{r_0}\right),\lb{potlog}\eqf and leads to a perturbing
radial acceleration \eqi  W_{r}=\rp{\gamma M}{r} \
.\lb{accel}\eqf In particular, in order to agree with the
potential (\ref{eq:sobo1}), we must set $\gamma=-\alpha/2$,
$r_0=2M$.

This kind of acceleration has been treated by \citet{SYR} with the approach outlined here getting negative answers.

\subsection{Yukawa-like corrections} \label{ssec:yuk}

In different works, both in the Palatini
\citep{Barraco93,Barraco96} and metric approach (e.g. see
\citep{Pech66,Stelle78,Capozziello07b})
Yukawa-like corrections are obtained. They lead to a gravitational
potential in the form

\beq \Phi(r)= -\frac{M}{r}\left[1+\alpha \exp \left(-\frac{r}{\lambda} \right) \right] \label{eq:yukawa1} \eeq

The parameter $\alpha$ is related to the strength of the correction, while $\lambda$ is related to the range of the modified potential.

The Yukawa correction to the Newtonian potential \eqi \Delta
\Phi(r)=-\rp{M\alpha}{r}\exp\left(-\rp{r}{\lambda}\right)\eqf
yields an entirely radial extra-acceleration \eqi W_r=-\rp{M\alpha}{r^2}\left(1+\rp{r}{\lambda}\right)\exp\left(-\rp{r}{\lambda}\right)
\lb{yacc}\eqf

By only assuming  $\lambda\gg ae$, i.e. Yukawa-type long-range modifications of gravity, it is possible to obtain useful approximated expressions
for the induced perihelion precession which, in turn,  allow to obtain  \citep{JHEP}
\eqi \lambda = \rp{a_{\rm B}-a_{\rm A}}{\ln\left(\sqrt{  \rp{a_{\rm B}}{a_{\rm A}}  } \rp{\Delta\dot\varpi_{\rm A}}{\Delta\dot\varpi_{\rm B}}\right)} \lb{ra_la}\eqf for the range
and
\eqi \alpha = \rp{2\lambda^2\Delta\dot\varpi}{\sqrt{Ma}}\exp\left(\rp{a}{\lambda}\right).\lb{yuk_al}\eqf
for the strength.
By using A $=$ Earth, B $=$ Mercury in \rfr{ra_la} one gets $\lambda = 0.182\pm 0.183$ AU; such a value for $\lambda$, the data of Venus and \rfr{yuk_al} yield $\alpha = ( -1\pm 4)\times 10^{-11}$ \citep{SYREXE}.

\subsection{Gravito-magnetic effects}\label{ssec:gm}

We have shown that the vacuum solutions of General Relativity with a cosmological
constant can be used in Palatini $f(R)$ gravity. In particular, the Kerr-de Sitter solution, which describes
a rotating black-hole in a space-time with a cosmological constant
\citep{demianski73,carter73,kerr03,kra04,kra05,kra07},  can be
used to investigate Gravito-magnetic effects in extended theories of gravity.

In particular, the weak-field and slow-motion approximation of the Kerr-de Sitter is \citep{iorioruggiero08}
\begin{align}
ds^2=&\left(1-\frac{2M}{r}-\frac{k }{3}r^2\right)dt^2 -
\left(1+\frac{2M}{r} -\frac{k }{6}r^2
\right)\left(dr^2+r^2 d\theta^2+r^2 \sin^2 \theta d\phi^2\right)+ \notag \\
& +2\frac{J}{M}\left(\frac{2M}{r}+\frac{k }{3}r^2+\frac{5}{6}
{M}{} k r  \right)\sin^2  \theta d\phi  dt.
\label{eq:kdsweak2}
\end{align}

We obtain the following expression for the perturbing gravito-magnetic potential
\eqI
\Delta A_{\varphi}= \frac{J}{M}\left(\frac{k }{3}r^2+\frac{5}{6}
{M}{} k r  \right)\sin^2  \theta.
\label{eq:deltaaphigm1}\eqF

Furthermore, the perturbing acceleration is
\eqI
\bds W= -2{{\bds v}}{}\bds\times \bcal{B}^{\rm G},\lb{accgm}\eqF where
the gravito-magnetic field $\bcal{B}^{\rm G}$ is  \eqI
\bcal{B}^{\rm G} = \rp{Jk}{3M}\bds{\hat{J}} +
\rp{5Jk}{12}\rp{\left[\bds{\hat{J}} +
\left(\bds{\hat{J}}\bds\cdot\bds{\hat{r}}\right)\bds{\hat{r}}\right]}{r}.\eqF

The resulting orbital effects are  \citep{iorioruggiero08}

\begin{eqnarray}\lb{orbi}
  \left\langle\dot a\right\rangle &=& 0, \\
  \left\langle\dot e\right\rangle &=& 0, \\
  \left\langle\dot i\right\rangle &=& 0, \\
  \left\langle\dot\Omega\right\rangle &=& \rp{Jk}{3M}\left(1+\rp{5M}{2 a}\right), \lb{k_O}\\
  \left\langle\dot\omega\right\rangle &=& -\rp{2Jk\cos i}{3M}\left(1 +\rp{5M}{4 a}\right),\lb{k_o} \\
  \left\langle\dot{\mathcal{M}}\right\rangle &=& {\overline{n}} + \rp{5Jk\cos i}{3M}\left(1 + \rp{M}{ a}\right).
\end{eqnarray}
In the calculation we have neglected terms of order $\mathcal{O}(e^2)$.

By using the corrections $\Delta\dot\varpi$ separately for each Solar System's planet one gets
$k\leq 10^{-29}$ m$^{-2}$.

For all the Solar System's planets the perihelion rate can be satisfactorily approximated by
\eqi \left\langle\dot\omega\right\rangle \approx -\rp{2Jk\cos i}{3M}.\eqf
Since $\cos i_{\rm A}/\cos i_{\rm B}\approx 1$ for every pair of planets A and B,
in this case $\mathcal{P}_{\rm A}/\mathcal{P}_{\rm B}\approx 1$; this possibility is ruled out by
$\mathcal{O}_{\rm A}/\mathcal{O}_{\rm B} = \Delta\dot\varpi_{\rm A}/\Delta\dot\varpi_{\rm B}$, as in the case of the DGP \citep{DGP} braneworld scenario \citep{AHEP}.

\section{Discussion and Conclusions} \label{sec:disconc}

In this paper we have considered some solutions of $f(R)$ gravity, both in the Palatini and metric formalism, that can be used to describe the weak gravitational field around the Sun. In particular, we have focused on the impact
that the modifications of the GR gravitational field, due to the non-linearity of $f(R)$,  have on the
Solar System dynamics. We have considered that these modifications have to be small in order not to contradict the known tests of GR and, as a consequence, we have treated them as perturbations. Thus, we have applied a perturbative approach to
compare the $f(R)$-induced  secular effects  with the latest observationally determinations  coming from various versions of the EPM planetary
ephemerides.  In particular, we have considered the ratios  of the
corrections to the standard secular precessions of the longitudes of perihelia estimated by E.V. Pitjeva for several pairs of planets in the Solar System. For all the models that we have considered (power law, Hooke-like force, logarithmic  corrections, Yukawa-like force, gravito-magnetic effects) our results show that the perturbations  deriving from the non-linearity of $f(R)$ are not compatible with the currently available  apsidal extra-precessions of the Solar System planets. Moreover, the hypothesis that the examined $f(R)$-induced perturbations   are zero, which cannot be tested by definition  with our approach, is  compatible with each perihelion extra-rate separately.

This might suggest that,  on the one hand,    the $f(R)$-induced secular effects cannot explain the observed extra-precessions and that, on the other hand, the effects of the non-linearity of the gravity Lagrangian are important on length scales much larger than the Solar System (e.g. on the cosmological scale) and  their effects on local physics are probably negligible.
It will be important to repeat such tests if and when other teams of astronomers will independently estimate their own corrections to the standard secular precessions of the perihelia.

\section*{Acknowledgments}

The authors would like to thank  Prof. P. M. Lavrov for his kind invitation to contribute to the  volume \textit{The Problems of Modern Cosmology},  on the occasion of the 50th birthday of Professor Sergei  D. Odintsov. L.I. gratefully thanks E.V. Pitjeva (Institute of Applied Astronomy, Russian Academy of Sciences, St. Petersburg) for useful and important information concerning her unpublished determinations of the perihelion precessions.
M.L.R. acknowledges financial support from the Italian Ministry of University and Research (MIUR) under the national program ``Cofin 2005'' - \textit{La
pulsar doppia e oltre: verso una nuova era della ricerca sulle pulsar}.


\end{document}